%% file: paper.tex
\documentclass[letterpaper,compsoc,twoside]{IEEEtran}
\usepackage{fixltx2e} % LaTeX patches, \textsubscript
\usepackage{cmap} % fix search and cut-and-paste in Acrobat
\usepackage{ifthen}
\usepackage[T1]{fontenc}
\usepackage[utf8]{inputenc}

\usepackage[font={small,it},labelfont=bf]{caption}
\usepackage{float}

\pdfoutput=1
\usepackage{scipy}
\makeatletter
\def\PY@reset{\let\PY@it=\relax \let\PY@bf=\relax%
    \let\PY@ul=\relax \let\PY@tc=\relax%
    \let\PY@bc=\relax \let\PY@ff=\relax}
\def\PY@tok#1{\csname PY@tok@#1\endcsname}
\def\PY@toks#1+{\ifx\relax#1\empty\else%
    \PY@tok{#1}\expandafter\PY@toks\fi}
\def\PY@do#1{\PY@bc{\PY@tc{\PY@ul{%
    \PY@it{\PY@bf{\PY@ff{#1}}}}}}}
\def\PY#1#2{\PY@reset\PY@toks#1+\relax+\PY@do{#2}}

\expandafter\def\csname PY@tok@gd\endcsname{\def\PY@tc##1{\textcolor[rgb]{0.63,0.00,0.00}{##1}}}
\expandafter\def\csname PY@tok@gu\endcsname{\let\PY@bf=\textbf\def\PY@tc##1{\textcolor[rgb]{0.50,0.00,0.50}{##1}}}
\expandafter\def\csname PY@tok@gt\endcsname{\def\PY@tc##1{\textcolor[rgb]{0.00,0.27,0.87}{##1}}}
\expandafter\def\csname PY@tok@gs\endcsname{\let\PY@bf=\textbf}
\expandafter\def\csname PY@tok@gr\endcsname{\def\PY@tc##1{\textcolor[rgb]{1.00,0.00,0.00}{##1}}}
\expandafter\def\csname PY@tok@cm\endcsname{\let\PY@it=\textit\def\PY@tc##1{\textcolor[rgb]{0.25,0.50,0.56}{##1}}}
\expandafter\def\csname PY@tok@vg\endcsname{\def\PY@tc##1{\textcolor[rgb]{0.73,0.38,0.84}{##1}}}
\expandafter\def\csname PY@tok@m\endcsname{\def\PY@tc##1{\textcolor[rgb]{0.13,0.50,0.31}{##1}}}
\expandafter\def\csname PY@tok@mh\endcsname{\def\PY@tc##1{\textcolor[rgb]{0.13,0.50,0.31}{##1}}}
\expandafter\def\csname PY@tok@cs\endcsname{\def\PY@tc##1{\textcolor[rgb]{0.25,0.50,0.56}{##1}}\def\PY@bc##1{\setlength{\fboxsep}{0pt}\colorbox[rgb]{1.00,0.94,0.94}{\strut ##1}}}
\expandafter\def\csname PY@tok@ge\endcsname{\let\PY@it=\textit}
\expandafter\def\csname PY@tok@vc\endcsname{\def\PY@tc##1{\textcolor[rgb]{0.73,0.38,0.84}{##1}}}
\expandafter\def\csname PY@tok@il\endcsname{\def\PY@tc##1{\textcolor[rgb]{0.13,0.50,0.31}{##1}}}
\expandafter\def\csname PY@tok@go\endcsname{\def\PY@tc##1{\textcolor[rgb]{0.20,0.20,0.20}{##1}}}
\expandafter\def\csname PY@tok@cp\endcsname{\def\PY@tc##1{\textcolor[rgb]{0.00,0.44,0.13}{##1}}}
\expandafter\def\csname PY@tok@gi\endcsname{\def\PY@tc##1{\textcolor[rgb]{0.00,0.63,0.00}{##1}}}
\expandafter\def\csname PY@tok@gh\endcsname{\let\PY@bf=\textbf\def\PY@tc##1{\textcolor[rgb]{0.00,0.00,0.50}{##1}}}
\expandafter\def\csname PY@tok@ni\endcsname{\let\PY@bf=\textbf\def\PY@tc##1{\textcolor[rgb]{0.84,0.33,0.22}{##1}}}
\expandafter\def\csname PY@tok@nl\endcsname{\let\PY@bf=\textbf\def\PY@tc##1{\textcolor[rgb]{0.00,0.13,0.44}{##1}}}
\expandafter\def\csname PY@tok@nn\endcsname{\let\PY@bf=\textbf\def\PY@tc##1{\textcolor[rgb]{0.05,0.52,0.71}{##1}}}
\expandafter\def\csname PY@tok@no\endcsname{\def\PY@tc##1{\textcolor[rgb]{0.38,0.68,0.84}{##1}}}
\expandafter\def\csname PY@tok@na\endcsname{\def\PY@tc##1{\textcolor[rgb]{0.25,0.44,0.63}{##1}}}
\expandafter\def\csname PY@tok@nb\endcsname{\def\PY@tc##1{\textcolor[rgb]{0.00,0.44,0.13}{##1}}}
\expandafter\def\csname PY@tok@nc\endcsname{\let\PY@bf=\textbf\def\PY@tc##1{\textcolor[rgb]{0.05,0.52,0.71}{##1}}}
\expandafter\def\csname PY@tok@nd\endcsname{\let\PY@bf=\textbf\def\PY@tc##1{\textcolor[rgb]{0.33,0.33,0.33}{##1}}}
\expandafter\def\csname PY@tok@ne\endcsname{\def\PY@tc##1{\textcolor[rgb]{0.00,0.44,0.13}{##1}}}
\expandafter\def\csname PY@tok@nf\endcsname{\def\PY@tc##1{\textcolor[rgb]{0.02,0.16,0.49}{##1}}}
\expandafter\def\csname PY@tok@si\endcsname{\let\PY@it=\textit\def\PY@tc##1{\textcolor[rgb]{0.44,0.63,0.82}{##1}}}
\expandafter\def\csname PY@tok@s2\endcsname{\def\PY@tc##1{\textcolor[rgb]{0.25,0.44,0.63}{##1}}}
\expandafter\def\csname PY@tok@vi\endcsname{\def\PY@tc##1{\textcolor[rgb]{0.73,0.38,0.84}{##1}}}
\expandafter\def\csname PY@tok@nt\endcsname{\let\PY@bf=\textbf\def\PY@tc##1{\textcolor[rgb]{0.02,0.16,0.45}{##1}}}
\expandafter\def\csname PY@tok@nv\endcsname{\def\PY@tc##1{\textcolor[rgb]{0.73,0.38,0.84}{##1}}}
\expandafter\def\csname PY@tok@s1\endcsname{\def\PY@tc##1{\textcolor[rgb]{0.25,0.44,0.63}{##1}}}
\expandafter\def\csname PY@tok@gp\endcsname{\let\PY@bf=\textbf\def\PY@tc##1{\textcolor[rgb]{0.78,0.36,0.04}{##1}}}
\expandafter\def\csname PY@tok@sh\endcsname{\def\PY@tc##1{\textcolor[rgb]{0.25,0.44,0.63}{##1}}}
\expandafter\def\csname PY@tok@ow\endcsname{\let\PY@bf=\textbf\def\PY@tc##1{\textcolor[rgb]{0.00,0.44,0.13}{##1}}}
\expandafter\def\csname PY@tok@sx\endcsname{\def\PY@tc##1{\textcolor[rgb]{0.78,0.36,0.04}{##1}}}
\expandafter\def\csname PY@tok@bp\endcsname{\def\PY@tc##1{\textcolor[rgb]{0.00,0.44,0.13}{##1}}}
\expandafter\def\csname PY@tok@c1\endcsname{\let\PY@it=\textit\def\PY@tc##1{\textcolor[rgb]{0.25,0.50,0.56}{##1}}}
\expandafter\def\csname PY@tok@kc\endcsname{\let\PY@bf=\textbf\def\PY@tc##1{\textcolor[rgb]{0.00,0.44,0.13}{##1}}}
\expandafter\def\csname PY@tok@c\endcsname{\let\PY@it=\textit\def\PY@tc##1{\textcolor[rgb]{0.25,0.50,0.56}{##1}}}
\expandafter\def\csname PY@tok@mf\endcsname{\def\PY@tc##1{\textcolor[rgb]{0.13,0.50,0.31}{##1}}}
\expandafter\def\csname PY@tok@err\endcsname{\def\PY@bc##1{\setlength{\fboxsep}{0pt}\fcolorbox[rgb]{1.00,0.00,0.00}{1,1,1}{\strut ##1}}}
\expandafter\def\csname PY@tok@kd\endcsname{\let\PY@bf=\textbf\def\PY@tc##1{\textcolor[rgb]{0.00,0.44,0.13}{##1}}}
\expandafter\def\csname PY@tok@ss\endcsname{\def\PY@tc##1{\textcolor[rgb]{0.32,0.47,0.09}{##1}}}
\expandafter\def\csname PY@tok@sr\endcsname{\def\PY@tc##1{\textcolor[rgb]{0.14,0.33,0.53}{##1}}}
\expandafter\def\csname PY@tok@mo\endcsname{\def\PY@tc##1{\textcolor[rgb]{0.13,0.50,0.31}{##1}}}
\expandafter\def\csname PY@tok@mi\endcsname{\def\PY@tc##1{\textcolor[rgb]{0.13,0.50,0.31}{##1}}}
\expandafter\def\csname PY@tok@kn\endcsname{\let\PY@bf=\textbf\def\PY@tc##1{\textcolor[rgb]{0.00,0.44,0.13}{##1}}}
\expandafter\def\csname PY@tok@o\endcsname{\def\PY@tc##1{\textcolor[rgb]{0.40,0.40,0.40}{##1}}}
\expandafter\def\csname PY@tok@kr\endcsname{\let\PY@bf=\textbf\def\PY@tc##1{\textcolor[rgb]{0.00,0.44,0.13}{##1}}}
\expandafter\def\csname PY@tok@s\endcsname{\def\PY@tc##1{\textcolor[rgb]{0.25,0.44,0.63}{##1}}}
\expandafter\def\csname PY@tok@kp\endcsname{\def\PY@tc##1{\textcolor[rgb]{0.00,0.44,0.13}{##1}}}
\expandafter\def\csname PY@tok@w\endcsname{\def\PY@tc##1{\textcolor[rgb]{0.73,0.73,0.73}{##1}}}
\expandafter\def\csname PY@tok@kt\endcsname{\def\PY@tc##1{\textcolor[rgb]{0.56,0.13,0.00}{##1}}}
\expandafter\def\csname PY@tok@sc\endcsname{\def\PY@tc##1{\textcolor[rgb]{0.25,0.44,0.63}{##1}}}
\expandafter\def\csname PY@tok@sb\endcsname{\def\PY@tc##1{\textcolor[rgb]{0.25,0.44,0.63}{##1}}}
\expandafter\def\csname PY@tok@k\endcsname{\let\PY@bf=\textbf\def\PY@tc##1{\textcolor[rgb]{0.00,0.44,0.13}{##1}}}
\expandafter\def\csname PY@tok@se\endcsname{\let\PY@bf=\textbf\def\PY@tc##1{\textcolor[rgb]{0.25,0.44,0.63}{##1}}}
\expandafter\def\csname PY@tok@sd\endcsname{\let\PY@it=\textit\def\PY@tc##1{\textcolor[rgb]{0.25,0.44,0.63}{##1}}}

\def\PYZus{\char`\_}

\def\PYZsh{\char`\#}

\def\PYZsq{\char`\'}

\makeatother

\providecommand*{\DUrole}[2]{%
  \ifcsname DUrole#1\endcsname%
    \csname DUrole#1\endcsname{#2}%
  \else% backwards compatibility: try \docutilsrole#1{#2}
    \ifcsname docutilsrole#1\endcsname%
      \csname docutilsrole#1\endcsname{#2}%
    \else%
      #2%
    \fi%
  \fi%
}

\ifthenelse{\isundefined{\hypersetup}}{
  \usepackage[colorlinks=true,linkcolor=blue,urlcolor=blue]{hyperref}
  \urlstyle{same} % normal text font (alternatives: tt, rm, sf)
}{}

\begin{document}
\newcounter{footnotecounter}\title{A Python-based Post-processing Toolset For Seismic Analyses}\author{Steve Brasier$^{\setcounter{footnotecounter}{1}\fnsymbol{footnotecounter}\setcounter{footnotecounter}{2}\fnsymbol{footnotecounter}}$%
          \setcounter{footnotecounter}{1}\thanks{\fnsymbol{footnotecounter} %
          Corresponding author: \protect\href{mailto:steve.brasier@atkinsglobal.com}{steve.brasier@atkinsglobal.com}}\setcounter{footnotecounter}{2}\thanks{\fnsymbol{footnotecounter} Atkins, 500 Park Avenue, Aztec West, BS32 4RZ}, Fred Pollard$^{\setcounter{footnotecounter}{2}\fnsymbol{footnotecounter}}$\thanks{%

          \noindent%
          Copyright\,\copyright\,2014 Steve Brasier et al. This is an open-access article distributed under the terms of the Creative Commons Attribution License, which permits unrestricted use, distribution, and reproduction in any medium, provided the original author and source are credited. http://creativecommons.org/licenses/by/3.0/%
        }}\maketitle
          \renewcommand{\leftmark}{PROC. OF THE 7th EUR. CONF. ON PYTHON IN SCIENCE (EUROSCIPY 2014)}
          \renewcommand{\rightmark}{A PYTHON-BASED POST-PROCESSING TOOLSET FOR SEISMIC ANALYSES}

\InputIfFileExists{page_numbers.tex}{}{}
\newcommand*{\docutilsroleref}{\ref}
\newcommand*{\docutilsrolelabel}{\label}
\AtEndDocument{\cleardoublepage}
\begin{abstract}This paper discusses the design and implementation of a Python-based
toolset to aid in assessing the response of the UK's Advanced Gas
Reactor nuclear power stations to earthquakes. The seismic analyses
themselves are carried out with a commercial Finite Element solver, but
understanding the raw model output this produces requires customised
post-processing and visualisation tools. Extending the existing tools had
become increasingly difficult and a decision was made to develop a new,
Python-based toolset. This comprises of a post-processing framework
(\texttt{aftershock}) which includes an embedded Python interpreter, and a
plotting package (\texttt{afterplot}) based on \texttt{numpy} and \texttt{matplotlib}.

The new toolset had to be significantly more flexible and easier to
maintain than the existing code-base, while allowing the majority of
development to be carried out by engineers with little training in software
development. The resulting architecture will be described with a focus on
exploring how the design drivers were met and the successes and challenges
arising from the choices made.\end{abstract}

\section{Introduction%
  \label{introduction}%
}

Nuclear power in the UK is provided by a fleet of Advanced Gas-cooled Reactors (AGRs) which became operational in the 1970's. These are a second generation reactor design and have a core consisting of layers of interlocking graphite bricks which act to slow neutrons from the fuel to sustain the fission reaction. Although the UK does not regularly experience significant earthquakes it is still necessary to demonstrate that the reactors could be safely shut-down if a severe earthquake were to occur.

The response of the graphite core to an earthquake is extremely complex and a series of computer models have been developed to simulate the behaviour. These models are regularly upgraded and extended as the cores change over their lives to ensure that the relevant behaviours are included. The models are analysed using the commercial Finite Element Analysis code LS-DYNA. This provides predicted positions and velocities for the thousands of graphite bricks in the core during the simulated earthquake.

By itself this raw model output is not particularly informative, and a complex set of post-processing calculations is required to help engineers to assess aspects such as:%
\begin{itemize}

\item 

Can the control rods still enter the core?
\item 

Is the integrity of the fuel maintained?
\end{itemize}

This post-processing converts the raw position and velocity data produced by the model into parameters describing the seismic performance of the core, assesses these parameters against acceptable limits, and presents the results in tabular or graphical form.

This paper describes a recent complete re-write of this post-processing toolset. It seeks to explore some of the software and architectural decisions made and examine the impact of these decisions on the engineering users.

\section{Background%
  \label{background}%
}

The LS-DYNA solver produces about 120GB of binary-format data for each simulation, split across multiple files. The existing post-processing tool was based on Microsoft Excel, using code written in Visual Basic for Applications (VBA) to decode the binary data and carry out the required calculations and Excel's graphing capabilities to plot the results. The original design of the VBA code was not particularly modular and its complexity had grown significantly as additional post-processing calculations were included and to accommodate developments in the models themselves. In short, there was significant \textquotedbl{}technical debt\textquotedbl{} \cite{Cun92} in the code which made it difficult to determine whether new functionality would adversely impact the existing calculations.

The start of a new analysis campaign forced a reappraisal of the existing approach as these issues meant there was low confidence that the new post-processing features required could be developed in the time or budget available. The following requirements were identified as strongly desirable in any new post-processing tool:%
\begin{itemize}

\item 

A far more modular and easily extensible architecture.
\item 

More flexible plotting capabilities.
\item 

A high-level, modern language to describe the actual post-processing calculations; these would be implemented by seismic engineers.
\item 

Better performance; the Excel/VBA post-processor could take 4-6 hours to complete which was inconvenient.
\item 

Possibility of moving to a Linux platform later, although starting initial development on Windows; this would allow post-processing to be carried out on a future Linux analysis server to streamline the work-flow and allow access to more powerful hardware.
\end{itemize}

A re-write from scratch would clearly be a major undertaking and was considered with some trepidation and refactoring the existing code would have been a more palatable first step. However further investigation convinced us that this would not progress a significant distance towards the above goals as the Excel/VBA platform was simply too limiting.

\section{Overall Architecture%
  \label{overall-architecture}%
}

An initial feasibility study lead to an architecture with three distinct parts:\newcounter{listcnt0}
\begin{list}{\arabic{listcnt0}.}
{
\usecounter{listcnt0}
\setlength{\rightmargin}{\leftmargin}
}

\item 

A central C++ core, \texttt{aftershock}, which handles the binary I/O and contains an embedded Python 2.7 interpreter.
\item 

A set of Python \textquotedbl{}calculation scripts\textquotedbl{} which define the actual post-processing calculations to be carried out.
\item 

A purpose-made Python plotting package \texttt{afterplot} which is based on \texttt{matplotlib} \cite{Hun07}.\end{list}

As the entire binary dataset is too large to fit in memory at once the \texttt{aftershock} core operates frame-by-frame, stepping time-wise through the data. At each frame it decodes the raw binary data and calls defined functions from the calculation scripts which have been loaded. These scripts access the data for the frame through a simple API provided by \texttt{aftershock} which returns lists of floats. The actual post-processing calculations defined by the scripts generally make heavy use of the \texttt{ndarrays} provided by \texttt{numpy} \cite{Wal11} to carry out efficient element-wise operations. As well as decoding the binary data and maintaining the necessary state for the scripts from frame-to-frame, the \texttt{aftershock} core also optimises the order in which the results files are processed to minimise the number of passes required.

The split between \texttt{afterplot} and a set of calculation scripts results in an architecture which:\setcounter{listcnt0}{0}
\begin{list}{\alph{listcnt0}.}
{
\usecounter{listcnt0}
\setlength{\rightmargin}{\leftmargin}
}

\item 

Has sufficient performance to handle large amounts of binary data.
\item 

Has a core which can be reused across all models and analyses.
\item 

Provides the required high-level language for \textquotedbl{}users\textquotedbl{}, i.e. the seismic engineers defining the calculations.
\item 

Hides the complex binary file-format entirely from the users.
\item 

Enforces modularity, separating the post-processing into individual scripts which cannot impact each other.\end{list}

With Python selected as the calculation scripting language a number of plotting packages immediately became options. However \texttt{matplotlib} \cite{Hun07} stood out for its wide use, \textquotedbl{}\emph{publication quality figures}\textquotedbl{} \cite{Hun07} and the sheer variety and flexibility of plotting capabilities it provided. Development of the post-processing toolset could have ended at this point, leaving the script engineers to utilise \texttt{matplotlib} directly. However \texttt{matplotlib}'s versatility comes with a price in complexity and the API is not particularly intuitive; requiring seismic engineers to learn the details of this did not seem to represent good value for the client. It was therefore decided to wrap \texttt{matplotlib} in a package \texttt{afterplot} to provide a custom set of very focussed plot formats.

\section{Plotting Architecture%
  \label{plotting-architecture}%
}

\texttt{afterplot} provides plotting functionality via a set of plotter classes, with the user (i.e. the engineer writing a calculation script) creating an instance of the appropriate class to generate a plot. All plotter classes inherit from a \texttt{BasePlot} class. This base class is essentially a wrapper for a \texttt{matplotlib} \texttt{Figure} object which represents a single plotting window, plus the \texttt{Axes} objects which represent the plots or sub-plots this contains.

At present \texttt{afterplot} provides only four types of plotter, although these are expected to be sufficient for most current requirements:\setcounter{listcnt0}{0}
\begin{list}{\arabic{listcnt0}.}
{
\usecounter{listcnt0}
\setlength{\rightmargin}{\leftmargin}
}

\item 

\texttt{LayerPlot} (Figure \DUrole{ref}{LayerPlot}): This represents values on a horizontal slice through the model using a contour-type plot but using discrete markers.
\item 

\texttt{ChannelPlot} (Figure \DUrole{ref}{ChannelPlot}): This represents the 3D geometry of a vertical column in the model by projection onto X-Z and Y-Z planes.
\item 

\texttt{TimePlot} (Figure \DUrole{ref}{TimePlot}): This is a conventional X-Y plot, representing time-histories as individual series with time on the X-axis.
\item 

\texttt{WaterfallPlot} (Figure \DUrole{ref}{WfallPlot}): This provides an overview of the distribution of the plotted parameter at each time-step during a simulation.\end{list}
\begin{figure*}[]\noindent\makebox[\textwidth][c]{\includegraphics[scale=0.60]{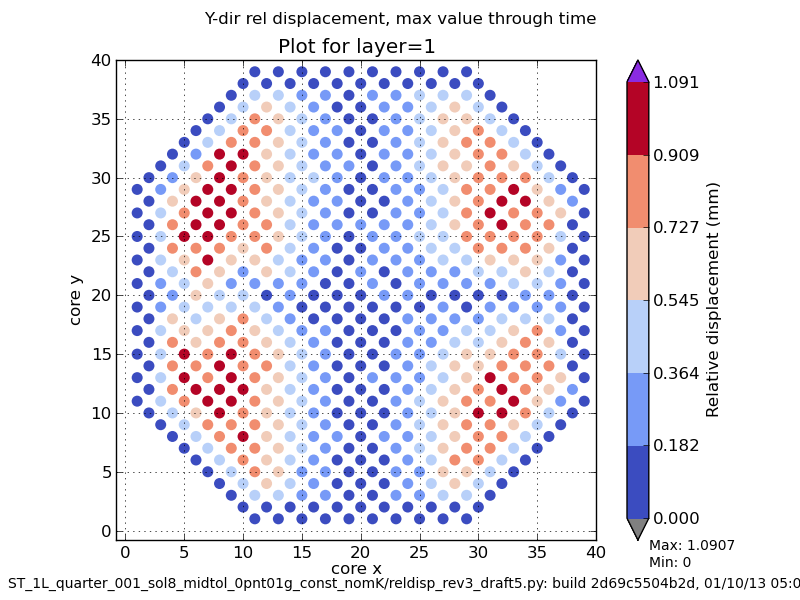}}
\caption{Example LayerPlot output \DUrole{label}{LayerPlot}}
\end{figure*}\begin{figure}[bht]\noindent\makebox[\columnwidth][c]{\includegraphics[scale=0.30]{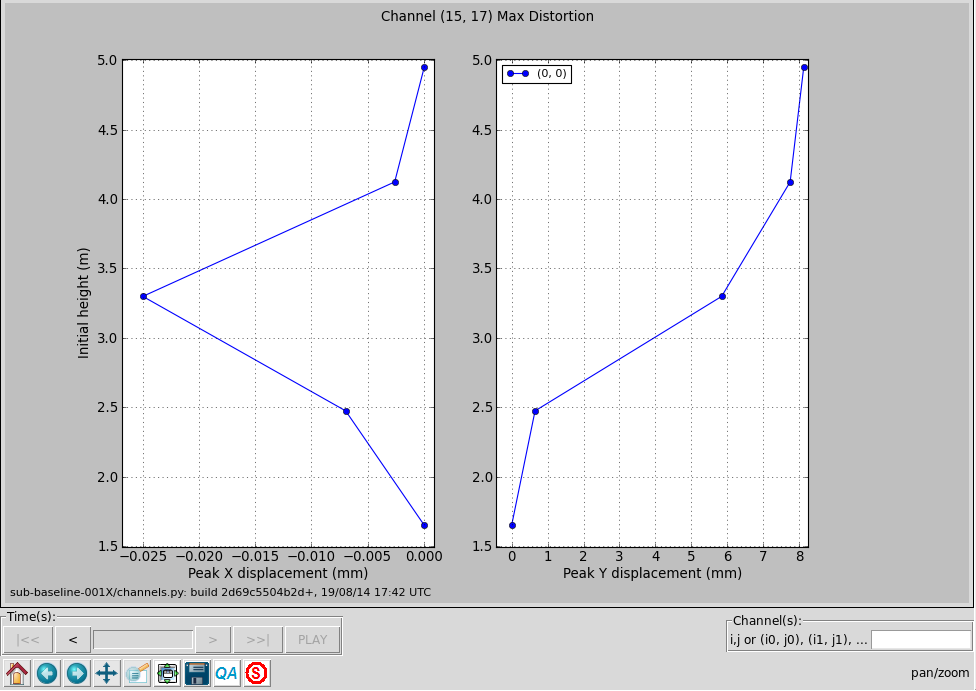}}
\caption{Example ChannelPlot with GUI \DUrole{label}{ChannelPlot}}
\end{figure}\begin{figure*}[]\noindent\makebox[\textwidth][c]{\includegraphics[scale=0.50]{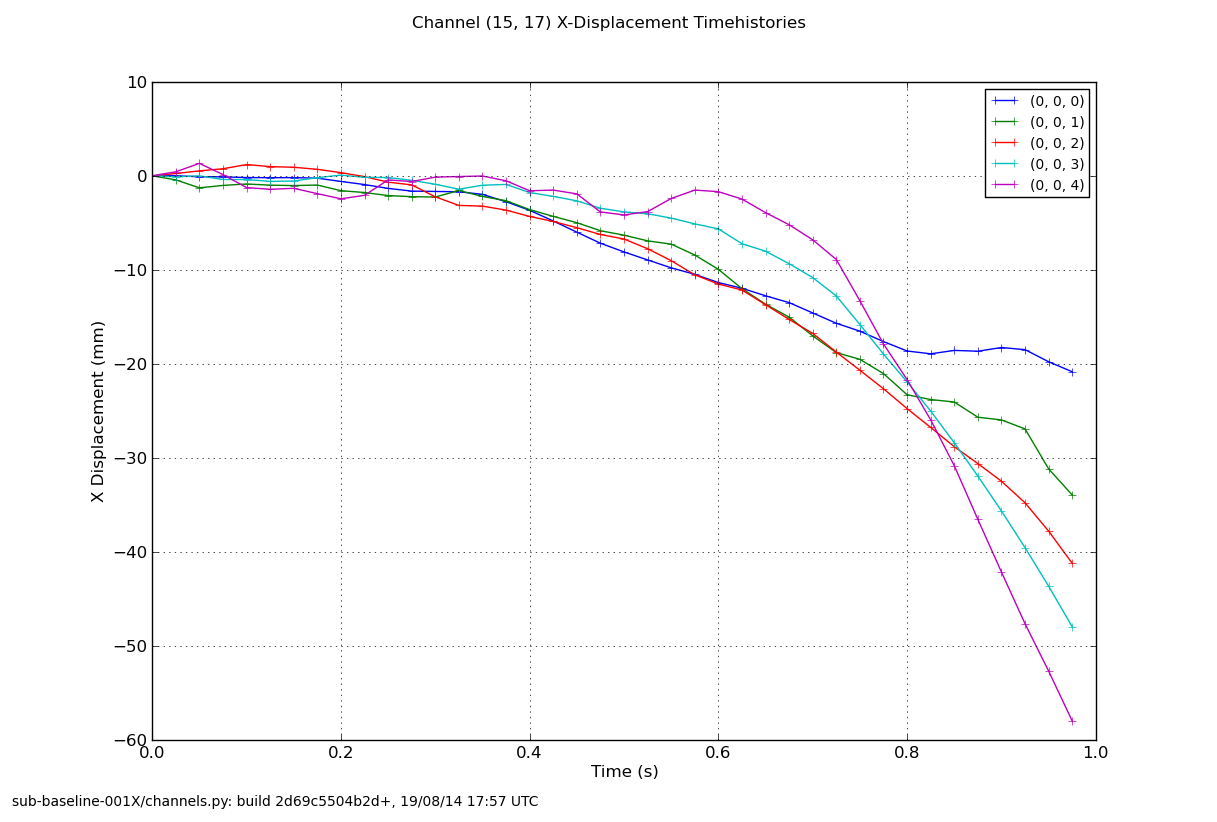}}
\caption{Example TimePlot output \DUrole{label}{TimePlot}}
\end{figure*}\begin{figure*}[]\noindent\makebox[\textwidth][c]{\includegraphics[scale=0.60]{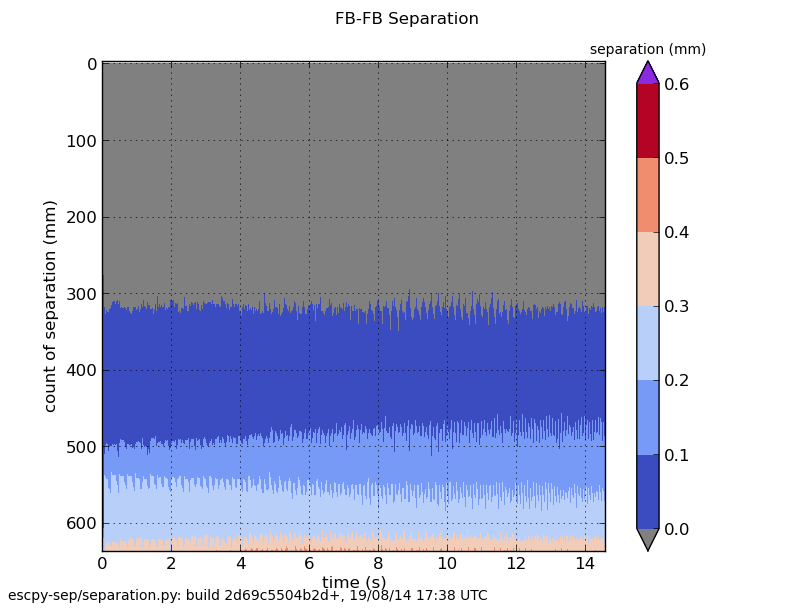}}
\caption{Example WaterfallPlot output \DUrole{label}{WfallPlot}}
\end{figure*}

Inherently all post-processed results are associated with a three-dimensional position within the model and a time within the simulation. Some parameters or outputs may collapse one or more of these dimensions, for example if plotting a plan view of peak values through time, maximums are taken over the vertical and time axes creating a set of results with two dimensions. All plotter classes therefore accept \texttt{numpy} arrays with up to four dimensions (or \texttt{axes} in numpy terminology). The meanings and order of these dimensions are standardised as three spatial dimensions followed by time, i.e. \texttt{(x, y, z, t)}, so that different \textquotedbl{}views\textquotedbl{} of the same data can easily be generated by passing an array to different plotters.

\section{Quality Advantages%
  \label{quality-advantages}%
}

A key advantage of providing a custom plotting package is that best-practice can be enforced on the generated plots, such as the provision of titles or use of grid-lines. Another example is that \texttt{afterplot} provides a custom   diverging colourmap as the default colourmap, based on the comprehensive discussion and methods presented in \cite{Mor09}. This should be significantly easier to interpret than the default colourmap provided by \texttt{matplotlib} in most cases.

The plotter classes can also allow \emph{alteration of presentation}, e.g. axis limits, while preventing \emph{modification of data}. Alteration of presentation is provided for by instance methods or GUI controls defined by the plotter classes. Modification of data is prevented simply by the lack of any interface to do this once the relevant array has been passed to the plot instance. This immutability is not intended as a security feature but simplifies quality assurance by limiting where errors can be introduced when altering presentation.

A further quality assurance feature is the capture of traceability data. When a new plot is generated, the \texttt{BasePlot} class traverses the stack frames using the standard library's \texttt{inspect} module to gather information about the paths and versions of calculation scripts and other Python modules used. This data is attached to the plots to assist in reproducing published plots or debugging issues. The use of introspection to capture this data means that this feature does not require any action by the script author.

\section{Interactive GUI%
  \label{interactive-gui}%
}

Providing a simple GUI was considered desirable to bridge the gap for users from the previous Excel-based toolset. The \texttt{matplotlib} documentation describes two methods of providing a GUI:\setcounter{listcnt0}{0}
\begin{list}{\arabic{listcnt0}.}
{
\usecounter{listcnt0}
\setlength{\rightmargin}{\leftmargin}
}

\item 

Using the cross-backend widgets provided in \texttt{matplotlib.widgets}, which are fairly limited.
\item 

Embedding the \texttt{matplotlib.FigureCanvas} object directly into the window provided by a specific GUI toolset such as \texttt{Tk}.\end{list}

An alternative approach is used by \texttt{afterplot} which is simpler than the second approach but allows the use of the richer widgets provided by specific GUI toolsets. This approach uses the \texttt{pyplot.figure()} function to handle all of the initial set-up of the GUI, with additional widgets then inserted using the GUI toolset's manager. This is demonstrated below by adding a \texttt{Tk} button to a \texttt{Figure} object using the \texttt{TkAgg} backend:\begin{Verbatim}[commandchars=\\\{\},fontsize=\footnotesize]
\PY{k+kn}{import} \PY{n+nn}{Tkinter} \PY{k+kn}{as} \PY{n+nn}{Tk}
\PY{k+kn}{import} \PY{n+nn}{matplotlib}
\PY{n}{matplotlib}\PY{o}{.}\PY{n}{use}\PY{p}{(}\PY{l+s}{\PYZsq{}}\PY{l+s}{TkAgg}\PY{l+s}{\PYZsq{}}\PY{p}{)}
\PY{k+kn}{from} \PY{n+nn}{matplotlib} \PY{k+kn}{import} \PY{n}{pyplot}
\PY{k}{class} \PY{n+nc}{Plotter}\PY{p}{(}\PY{n+nb}{object}\PY{p}{)}\PY{p}{:}
  \PY{k}{def} \PY{n+nf}{\PYZus{}init\PYZus{}\PYZus{}}\PY{p}{(}\PY{n+nb+bp}{self}\PY{p}{)}\PY{p}{:}
    \PY{n+nb+bp}{self}\PY{o}{.}\PY{n}{figure} \PY{o}{=} \PY{n}{pyplot}\PY{o}{.}\PY{n}{figure}\PY{p}{(}\PY{p}{)}
    \PY{n}{window} \PY{o}{=} \PY{n+nb+bp}{self}\PY{o}{.}\PY{n}{figure}\PY{o}{.}\PY{n}{canvas}\PY{o}{.}\PY{n}{manager}\PY{o}{.}\PY{n}{window}
    \PY{n}{btn\PYZus{}next} \PY{o}{=} \PY{n}{Tk}\PY{o}{.}\PY{n}{Button}\PY{p}{(}\PY{n}{master}\PY{o}{=}\PY{n}{window}\PY{p}{,}
                         \PY{n}{text}\PY{o}{=}\PY{l+s}{\PYZsq{}}\PY{l+s}{next}\PY{l+s}{\PYZsq{}}\PY{p}{,}
                         \PY{n}{command}\PY{o}{=}\PY{n+nb+bp}{self}\PY{o}{.}\PY{n}{\PYZus{}next}\PY{p}{)}
    \PY{n}{btn\PYZus{}next}\PY{o}{.}\PY{n}{pack}\PY{p}{(}\PY{n}{side}\PY{o}{=}\PY{n}{Tk}\PY{o}{.}\PY{n}{LEFT}\PY{p}{)}
    \PY{n+nb+bp}{self}\PY{o}{.}\PY{n}{figure}\PY{o}{.}\PY{n}{show}\PY{p}{(}\PY{p}{)}
\end{Verbatim}

\section{Store and Restore%
  \label{store-and-restore}%
}
Functionality to save plots to disk as images is provided by \texttt{matplotlib} via \texttt{Figure.savefig()} which can generate a variety of formats. When development of \texttt{afterplot} began a \texttt{matplotlib.Figure} object could not be pickled and therefore there was no native way to regenerate it for interactive use except for re-running the script which created it. Despite the improved performance provided by \texttt{aftershock} this is clearly time-consuming when only minor presentation changes are required such as altering the limits on an axis. A means to enable an entire plotter instance, including its GUI, to be stored to disk and later restored to a new fully interactive GUI was therefore strongly desirable. While the ability to pickle \texttt{Figure} objects has since been added to \texttt{matplotlib} this would not support the custom GUIs which \texttt{afterplot} provides. However, by following the same approach that the \texttt{pickle} module uses internally to handle class instances the desired store/restore functionality could be added relatively simply.

\textbf{Storing:}\setcounter{listcnt0}{0}
\begin{list}{\arabic{listcnt0}.}
{
\usecounter{listcnt0}
\setlength{\rightmargin}{\leftmargin}
}

\item 

When a plot instance is created, the \texttt{\_\_new\_\_} method of the \texttt{BasePlot} superclass binds the  supplied \texttt{*args} and \texttt{**kwargs} to attributes on the plotter instance - these will include one or more \texttt{ndarrays} containing the actual data to be plotted.
\item 

To store the instance, first a \texttt{type} object is obtained, then this and the \texttt{*args} and \texttt{**kwargs} are pickled.\end{list}

Simplified code for the \texttt{BasePlot} class implementing storing to a given path:\begin{Verbatim}[commandchars=\\\{\},fontsize=\footnotesize]
\PY{k}{class} \PY{n+nc}{BasePlot}\PY{p}{(}\PY{n+nb}{object}\PY{p}{)}\PY{p}{:}
  \PY{k}{def} \PY{n+nf}{\PYZus{}\PYZus{}new\PYZus{}\PYZus{}}\PY{p}{(}\PY{n}{cls}\PY{p}{,} \PY{o}{*}\PY{n}{args}\PY{p}{,} \PY{o}{*}\PY{o}{*}\PY{n}{kwargs}\PY{p}{)}\PY{p}{:}
    \PY{n}{obj} \PY{o}{=} \PY{n+nb}{object}\PY{o}{.}\PY{n}{\PYZus{}\PYZus{}new\PYZus{}\PYZus{}}\PY{p}{(}\PY{n}{cls}\PY{p}{)}
    \PY{n}{obj}\PY{o}{.}\PY{n}{\PYZus{}args}\PY{p}{,} \PY{n}{obj}\PY{o}{.}\PY{n}{\PYZus{}kwargs} \PY{o}{=} \PY{n}{args}\PY{p}{,} \PY{n}{kwargs}
    \PY{k}{return} \PY{n}{obj}
  \PY{k}{def} \PY{n+nf}{store}\PY{p}{(}\PY{n+nb+bp}{self}\PY{p}{,} \PY{n}{path}\PY{p}{)}\PY{p}{:}
    \PY{n}{data} \PY{o}{=} \PY{p}{(}\PY{n+nb}{type}\PY{p}{(}\PY{n+nb+bp}{self}\PY{p}{)}\PY{p}{,} \PY{n+nb+bp}{self}\PY{o}{.}\PY{n}{\PYZus{}args}\PY{p}{,} \PY{n+nb+bp}{self}\PY{o}{.}\PY{n}{\PYZus{}kwargs}\PY{p}{)}
    \PY{k}{with} \PY{n+nb}{open}\PY{p}{(}\PY{n}{path}\PY{p}{,} \PY{l+s}{\PYZsq{}}\PY{l+s}{w}\PY{l+s}{\PYZsq{}}\PY{p}{)} \PY{k}{as} \PY{n}{pkl}\PY{p}{:}
      \PY{n}{pickle}\PY{o}{.}\PY{n}{dump}\PY{p}{(}\PY{n}{data}\PY{p}{,} \PY{n}{pkl}\PY{p}{)}
    \PY{k}{def} \PY{n+nf}{show}\PY{p}{(}\PY{n+nb+bp}{self}\PY{p}{)}\PY{p}{:}
      \PY{c}{\PYZsh{} .. gui code here ..}
\end{Verbatim}
\textbf{Restoring}:\setcounter{listcnt0}{0}
\begin{list}{\arabic{listcnt0}.}
{
\usecounter{listcnt0}
\setlength{\rightmargin}{\leftmargin}
}

\item 

The type object, \texttt{args} and \texttt{kwargs} are unpickled from the file.
\item 

The type object is called to create a new instance, passing it the unpickled \texttt{args} and \texttt{kwargs}.\end{list}

Simplified restoring code, taking a path to a stored file and regenerating the plot complete with interactive GUI:\begin{Verbatim}[commandchars=\\\{\},fontsize=\footnotesize]
\PY{k}{def} \PY{n+nf}{restore}\PY{p}{(}\PY{n}{path}\PY{p}{)}\PY{p}{:}
  \PY{k}{with} \PY{n+nb}{open}\PY{p}{(}\PY{n}{path}\PY{p}{,} \PY{l+s}{\PYZsq{}}\PY{l+s}{r}\PY{l+s}{\PYZsq{}}\PY{p}{)} \PY{k}{as} \PY{n}{pkl}\PY{p}{:}
    \PY{n}{t\PYZus{}plt}\PY{p}{,} \PY{n}{args}\PY{p}{,} \PY{n}{kwargs} \PY{o}{=} \PY{n}{pickle}\PY{o}{.}\PY{n}{load}\PY{p}{(}\PY{n}{pkl}\PY{p}{)}
    \PY{n}{restored\PYZus{}plotter} \PY{o}{=} \PY{n}{t\PYZus{}plt}\PY{p}{(}\PY{o}{*}\PY{n}{args}\PY{p}{,} \PY{o}{*}\PY{o}{*}\PY{n}{kwargs}\PY{p}{)}
    \PY{n}{restored\PYZus{}plotter}\PY{o}{.}\PY{n}{show}\PY{p}{(}\PY{p}{)}
\end{Verbatim}
Note that classes can define \texttt{\_\_getstate\_\_} and \texttt{\_\_setstate\_\_} methods to control how they are pickled and un-pickled and the approach described above could be implemented in these methods. The use of explicitly-named methods and functions was primarily to make the approach transparent to future developers.

This approach has a number of benefits:\setcounter{listcnt0}{0}
\begin{list}{\arabic{listcnt0}.}
{
\usecounter{listcnt0}
\setlength{\rightmargin}{\leftmargin}
}

\item 

Neither the storing nor restoring code needs to know anything about the actual plot class, except that it has a \texttt{show()} method, hence any plotter derived from \texttt{BasePlot} inherits this functionality.
\item 

The only interface which storing and restoring needs to address is the plotter class's signature. This is simple and robust, as code can always be added to a class's \texttt{\_\_init\_\_} method to handle changes in the signature such as depreciated parameters, meaning that it should essentially always be possible to make stored plots forward-compatible with later versions of \texttt{afterplot}. By contrast normally when a class instance is unpickled, \texttt{pickle} directly sets the instance's dictionary (the \texttt{\_\_dict\_\_} attribute) from the pickled data, meaning that changes to an instance's internal attributes can break unpickling.
\item 

Restoring the interactive GUI requires no additional code - only what is needed to create the GUI when the plotter instance is first created.
\item 

If a stored plot is restored with a later version of \texttt{afterplot} any enhanced GUI functionality will automatically be available.\end{list}

For convenience a simple \texttt{cmd} script and short Python function also allow stored plots to be restored on user's local Windows PCs and the GUI displayed by simply double-clicking the file. Alternatively a simple script can be written to batch process presentational changes such as colour bars or line thicknesses for a series of plots. Such a script uses a provided \texttt{restore()} function to restore the desired plots without showing the GUI, then uses the methods the plotter classes provide to alter desired presentation aspects.

One complication omitted from the simplified code above is that ideally storing and restoring should be insensitive to whether parameters have been specified as positional or named arguments. Therefore the \texttt{\_\_new\_\_()} method of the \texttt{BasePlot} superclass uses \texttt{inspect.getargspec()} to convert all arguments to a dictionary of \texttt{name:value}. Class instances are then actually stored/restored as if all parameters were provided as keyword arguments.

While this approach essentially mirrors how \texttt{pickle} handles class instances, implementing such complex and robust
functionality in such little code is an impressive demonstration of Python’s benefits.

\section{Outcomes and Lessons Learnt%
  \label{outcomes-and-lessons-learnt}%
}

The overall architecture has been a success:%
\begin{itemize}

\item 

Performance is significantly improved.
\item 

Post-processing can easily be integrated with analysis runs if required.
\item 

Maintainability and extensibility of the calculations has been vastly improved.
\item 

Python and \texttt{numpy} form a vastly more usable and concise high-language for describing calculations than VBA, allowing engineers to concentrate on the logic rather than working around the language.
\item 

The \texttt{aftershock} core is reusable across different models which will save considerable effort for future models.
\item 

Cross-platform portability to Windows and Linux was achieved without any significant effort for the calculation scripts and plotting module, making a decision to transition part-way through the project to new Linux hardware relatively straightforward.
\end{itemize}

However there were a number of challenges, some of which were expected at the outset and some which were not:

\emph{Education and training:} As discussed a key driver for the architecture was that the calculation scripts would be written by seismic engineers, as they were the domain experts. Some of these engineers were already familiar with Python, often from scripting environments provided by commercial analysis software, or with other high-level scripting languages such as VBA. In general users found it relatively simple to pick up and start developing procedural and simple object-orientated Python, but the heavy use of \texttt{numpy} for element-wise operations then required users to learn a third programming paradigm. While the basic concepts were easily understood, deciding when the use of explicit loops or element-wise operations is more appropriate requires considerably more experience. Most engineers had not written code where performance was a concern and hence basic optimisation techniques such as moving constant expressions outside of loops were not necessarily considered obvious. Inconsistencies in the API for the scientific Python stack also led to some subtle performance and functionality issues; for example the three examples below all have different answers as to which package is \textquotedbl{}best\textquotedbl{}:%
\begin{itemize}

\item 

\texttt{abs()} vs. \texttt{numpy.abs()}
\item 

\texttt{math.exp()} vs. \texttt{numpy.exp()},
\item 

\texttt{math.pi} vs. \texttt{scipy.pi} vs. \texttt{numpy.pi}
\end{itemize}

\emph{Development practicalities}: Some significant difficulties were encountered in compiling \texttt{afterplot} on both Windows and Linux due to the embedded Python 2.7 interpreter, but these issues are outside the scope of this paper to discuss.

\emph{Plotting functionality:} The success of the \texttt{afterplot} plotting module is less clear at present. It has provided the desired plotting flexibility, as demonstrated by the \texttt{LayerPlot} and \texttt{WaterfallPlot} plot types which could not be easily replicated using Excel's plotting facilities. The control of style it enforces also appears to be strongly desirable in terms of reducing the effort required to obtain publication-quality plots. However verification of the relatively complex GUI code has proved to be difficult. \textquotedbl{}Verification\textquotedbl{} in this sense does not refer to a formal proof of correctness, but to a level of independent checking consistent with that applied to the actual post-processing calculations. Part of the difficulty with this was due to the limited internal availability of developers familiar with the GUI toolset. Another aspect was the decision to provide a small number of relatively general-purpose plot classes. This made it necessary for the plot classes to accept data in different dimensions and with a variety of options, complicating the internal logic which often involves complex array striding and reshaping. It may have been simpler overall to provide a larger number of less flexible plotters with simpler interfaces and fewer internal code paths. Testing plotting code is not straightforward but \texttt{matplotlib}'s own test suite has provided some useful techniques to automatically check images produced by test cases against known-good results.

Overall, the decision to use the Python scientific software stack for this toolset has been strongly positive. Encouragingly it also appears that future develoments are likely to provide features like sparse arrays and lazy evaluation which would permit the calculation scripts to be simpler and more efficient. Similarly, rationalisation of the \texttt{matplotlib} API is expected in future which will simplify the creation of high-quality plots from Python.

\end{document}

%% file: page_numbers.tex
\setcounter{page}{73}